\documentclass[preprint,sort,number,12pt]{elsarticle}
\usepackage{amssymb}
\usepackage{epsfig}
\usepackage{colordvi}
\journal{Nuclear Physics A}
\begin{document}
\begin{frontmatter}

\title{
Constraining the nuclear matter equation of state around twice
saturation density 
}
\author[label1]{A.~Le F\`evre}
\author[label1]{Y.~Leifels}
\author[label1]{W.~Reisdorf}
\author[label2]{J.~Aichelin}
\author[label2]{Ch.~Hartnack}
\address[label1]{GSI Helmholtzzentrum f\"ur Schwerionenforschung GmbH,
 Darmstadt, Germany}
\address[label2]{SUBATECH, UMR 6457, Ecole des Mines de Nantes - IN2P3/CNRS - Universit\'e de Nantes, France}



\begin{abstract}
Using FOPI data on elliptic flow in Au+Au collisions between 0.4 and 1.5A~GeV 
we extract constraints for the 
equation of state (EOS) of compressed symmetric nuclear matter 
using the transport code IQMD 
by introducing an observable describing the evolution 
of the size of the elliptic flow as a function of rapidity. This observable 
is sensitive to the nuclear EOS and a robust tool to constrain the compressibility
of nuclear matter up to 2 $\rho_0$.

\end{abstract}

\end{frontmatter}


\section{Introduction}\label{Introduction}

The equation of state (EOS) of nuclear matter is of fundamental interest
and has been the object of intense theoretical efforts since several
decades. 
The interest is boosted by the fact that it is an important ingredient
in modeling fascinating astrophysical phenomena such as compact stars
\cite{lattimer12}
and core collapse supernovae \cite{burrows13}.
The calculation of the nuclear EOS from first principles, such as very recently
attempted in \cite{gezerlis13}, is a very complex task.

The  theory of nuclear forces,
such as reviewed in \cite{machleidt11},
requires a confrontation with empirical facts, when it is applied to the
'ab initio' problem of EOS calculations.
In the last decade astrophysicists have pointed out (for a recent review, see
\cite{lattimer12}) that the EOS relevant for compact stars can in principle be
deduced from the increasingly accurate and voluminous systematics of
'neutron' star masses and radii 
by use of the relativistic 
Tolman-Volkov-Oppenheimer (TOV) equations.
However, at the present time precise model-independent radii are still
missing \cite{lattimer14}, while on the mass front line 
considerable progress has been made
\cite{demorest10,antoniadis13}.
Nevertheless the need of additional information remains: 
the composition of matter in
the core of neutron stars is not known so far.

A complementary method to approach the nuclear equation of state, 
practiced since the mid-eighties, is the use of heavy
ion collisions over a wide range of incident energies, system sizes and compositions.
The general difference between astrophysical and earth bound experiments is
that the latter ones deal with rather isospin symmetric ($N\approx Z$, $N$ neutron,
$Z$ proton number) system compositions than are expected in the center of compact
stars, and also, presumably, the temperatures in such reactions are much
higher.  
There are, besides astrophysics, very good reasons to study symmetric (or
almost symmetric) nuclear matter.
It is clear that we must understand symmetric nuclear matter if we want
to explain the properties of nuclei.
Convincing theories  are expected to reproduce properties
of symmetric and 'pure' (or 'almost pure') neutron matter on the same
footing.
Also the possibility of a phase transition as the density/temperature
is raised can only be studied quantitatively in heavy ion reactions.
  
However, deriving  constraints on the nuclear equation of state (EOS) from
heavy-ion collision data is a highly non-trivial
task: we are in a dynamic situation with the characteristics describing the
state of the system varying rapidly in time, while the
observables are measured long after the various incoming and created parts 
of the system have separated forming a three-dimensional momentum pattern
filling large (but not all) parts of phase space.
Clearly, the claim to a significant 'constraint' on the EOS cannot
easily be made 'model-independent'. 

So far attempts to derive reasonably narrow constraints are limited to
the energy regime below $10A$ GeV.
The reason is that exclusively for this energy regime some kind of a clock is
available. 
As described in a short, but pedagogical description \cite{danielewicz02}
of the reaction scenario in this regime, 
a clock is provided by the fact that 
the 'spectator' nucleons, not initially hit in collisions with non-zero
impact parameters, are in contact with  the more central 'participant'
matter that forms the compressed 'fireball' for a defined time,
 the passing time, or ``blockage time'' \cite{danielewicz02}.
This time  turns out to be comparable to the expansion time.
The latter is governed by a fundamental property, the sound velocity
which connects to the EOS.

The price one has to pay for a 'clock' consists in
more complex shapes \cite{clare86,hombach99,danielewicz02} due to the presence
of spectators. 
In contrast, at energies beyond 10A GeV one can essentially start from
'simple' almond or elliptic shaped participant clusters of nucleons whose
evolution generating dominantly in plane 'elliptic' flow was considered 
by Ollitrault \cite{ollitrault92} in the early nineties.
However the spectator clock is no longer operative.


But even in this (nowadays 'low to intermediate') energy regime the
present knowledge on the properties of compressed nuclear matter is not satisfactory.

Hopes that pion yields, measured in the mid-eighties, would be a textbook
signal of a stiff EOS \cite{stock86} have not materialized in the sequel.
For possibly related reasons, the 
interpretation of pion isospin {\it ratio} data in helping to
constrain the EOS of asymmetric systems is not settled 
\cite{xiao09,zqfeng10}.

The use of positively charged kaons, $K^+$, in the subthreshold region
proposed by \cite{aichelin85} has been more successful.
It is based on the principle that $K^+$ are produced in larger amounts
if the nuclear EOS is soft due to the higher achieved density causing
more frequent collisions. 
This effect is enhanced in the deeply subthreshold energy regime because the
production of $K^+$ then requires multi-step reactions, including $\Delta$
production, in the medium.
In addition, the relatively low reabsorption cross sections are
supposed to favor a good memory of the high density phase.

Using data from the KAOS Collaboration on heavy ion systems of varying
size (namely C+C and Au+Au) and hence achieved
density \cite{sturm01}, two theoretical groups \cite{fuchs,hartnack}
concluded that  the so-called 'soft' EOS (see later) was best suited to
explain the data.
Despite this encouraging success some questions remain open.
The kaon is very rare at the lowest measured incident energy, $0.8A$ GeV,
where the sensitivity to the nuclear EOS is largest and one has to ensure
that all 'bulk' observables (multiplicities, clusterization, stopping, flow)
are under control.
On the experimental side some desirable data, needed to
complete the picture, are still missing: 
when switching from carbon to gold one is varying, besides the size, also
the proton-neutron asymmetry, but the isospin partner, $K^0$, has not been
measured at $0.8A$ GeV.
Also, detailed experimental data on the nucleonic observables in C+C, 
needed to corroborate the simulated densities and momentum dependencies, 
are not available.

Clearly other observables such as particle flow are welcome in order
to confirm the conclusions from the 'kaon method'. 

In a study \cite{swang96} of Au+Au collisions covering the beam energy range
of $0.25A$ to $1.15A$ GeV the EoS Collaboration made an attempt to
establish a 'complete characterization of squeeze-out' (preferential emission
out of the reaction plane) including a simultaneous consideration of
'three categories of collective motion' (radial and sideward flow and
squeeze-out).
From the work which was focused on the $0.6A$ GeV data, the authors
concluded at the time (1996) that 'the QMD (a transport code \cite{peilert89})
comparisons do not show a level of agreement with the data that justifies
inferences about nuclear incompressibility'.

Almost 10 years later (2005) the FOPI Collaboration published 
\cite{andronic05} an excitation
function of 'elliptic flow' (see later) extending from $0.09A$ to $1.5A$ GeV
beam energy for $Z=1$ particles, also in the Au+Au system.
The data were compared to the predictions from four different transport codes.
The conclusion was 'no strong constraint on the EOS can be derived at this
stage'.

A transport theoretical study of  sideflow and elliptic flow data
(restricted to emitted protons) both
from the BEVALAC and the AGS  accelerators
was undertaken in \cite{danielewicz02}.
The authors characterized their 'trial' EOS by the incompressibility $K_0$ of
of symmetric nuclear matter around saturation density 
($\rho=\rho_0\approx 0.16$ fm$^{-3}$).
An EOS labeled by $K_0=167$ MeV was adjusted
to the experimental data on directed flow at incident energies below 1$A$ Gev,
with a better description of the higher energy data using $K_0=200$ MeV.
The elliptic flow data seemed to require $K_0=300$ MeV.

In the present work we intend to improve the situation in the $1A$ GeV regime.
Profitting from extensive flow data published recently by the FOPI
Collaboration \cite{reisdorf12} we take here specifically a close look at
the elliptic flow data, taking advantage of the afore-mentioned clock.
In doing so 

1. we use not only protons, but also two- and three-nucleon clusters
that are emitted in Au+Au collisions in the incident beam energy range
from $0.4A$ to $1.5A$ GeV and that have larger flow signals than
single nucleons

2. we use not only mid-rapidity data, but use the information from
$80\%$ of the target-projectile rapidity gap; as we will show this information
can be lumped into a single observable.

The data are confronted to a well established transport code,
Isospin Quantum Molecular Dynamics, IQMD \cite{aichelin91,hartnack98}
using various phenomenological
EOS to assess the EOS that is most compatible with the FOPI data.
Four Skyrme type parameterizations 
are available in the code: $H$ ('stiff'), $S$ ('soft'),
$HM$ ('stiff momentum dependent'), $SM$ ('soft momentum dependent').
As in \cite{danielewicz02} we characterize the EOS (for details see
\cite{hartnack98}) by saturation density incompressibilities: 
$K_0=200$ ('soft') and $K_0=380$ ('stiff').

We also estimate an error band from uncertainties in the data and compare
to some recent representative microscopic EOS calculations.

In the following section we present the analysis of the elliptic flow data
obtained by FOPI \cite{reisdorf12} and the resulting FOPI-IQMD constraint on the
EOS. 
Dynamical details from the IQMD simulations are presented in section 
\ref{Simulations} and the 
constraints for the EOS obtained from the FOPI to IQMD comparisons 
are confronted with representative theoretical calculations in section
\ref{Comparison}.
Finally, in section \ref{Discussion} we briefly discuss the obtained results
and the needs for future efforts. 
  
\section{Analysis and results}\label{Analysis}

A comprehensive documentation on flow data 
in the SIS energy range
measured by the FOPI Collaboration is available in \cite{reisdorf12}.
We shall therefore be brief here.

Owing to collective flow phenomena, discovered experimentally in 
1984~\cite{gustafsson84,renfordt84},
it is possible to reconstruct the reaction plane 
event-by-event and hence to study
azimuthal correlations relative to that plane.
The well established parameterization \cite{voloshin96,poskanzer98}
of the observed azimuthal, $\phi$, distributions is used:
\begin{equation}
 u = (\gamma , \vec{\beta}\gamma) \ ;  \ \  u_{t} = \beta_{t}\gamma 
\end{equation}
\begin{equation}
 \frac{dN}{u_t du_t dy d\phi} = v_0 [1 + 2v_1 \cos(\phi) + 2v_2
\cos(2\phi)] 
\end{equation}
\begin{equation}
 v_0 = v_0(y,u_t) \  ; \ \  v_1 = v_1(y,u_t) \ ; \ \  v_2 = v_2(y,u_t) 
\end{equation}
\begin{equation}
 v_1 = \left<\frac{u_x}{u_t}\right> = \left <cos(\phi) \right >\ ;
 \ \  v_2 = \left <\left( {\frac{u_x}{u_t}}\right)^2 -
  {\left(\frac{u_y}{u_t}\right)}^2 \right > = \left <cos(2\phi)\right >
\end{equation}

Here we introduced the longitudinal (beam axis) rapidity $y$
in the  ($c.o.m.$) reference system and  the transverse
(spatial) component $t$ of the four-velocity $u$, given by $u_t=\beta_t\gamma$.
The 3-vector $\vec{\beta}$ is the velocity in units of the light
velocity and $\gamma=1/\sqrt{1-\beta^2}$.
Throughout we use scaled units $y_0=y/y_p$ and $u_{t0}=u_t/u_p$,
with $u_p=\beta_p \gamma_p$, the index p referring to the incident projectile
in the $c.o.m.$.
In these units the initial target-projectile rapidity gap always extends from
$y_0=-1$ to $y_0=1$.
In \cite{reisdorf12} the transverse momentum method~\cite{danielewicz85}
was used including all particles
identified outside the mid-rapidity interval $|y_0|<0.3$ and excluding
identified pions.

\begin{figure}[t]
\centering
\includegraphics[width=0.70\textwidth]{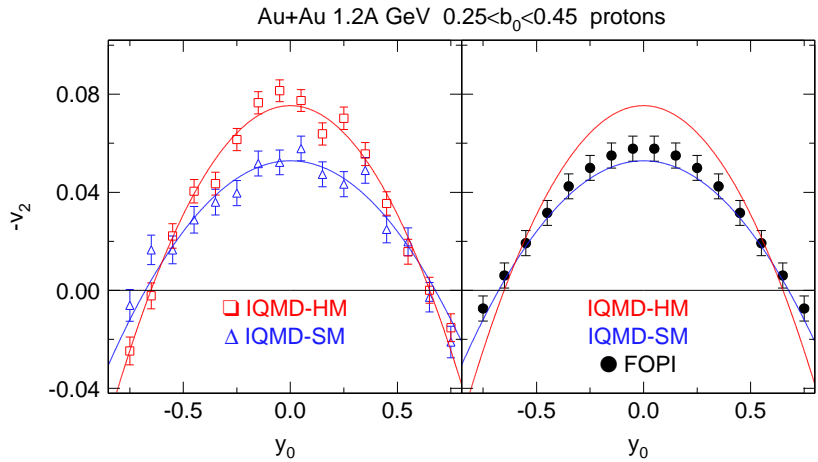}
\caption{Proton elliptic flow data, $-v_2(y_0)$, and IQMD-SM/HM simulations.
See text for further explanations.}
\label{v2}
\end{figure}

In Fig.~\ref{v2} we show a sample of proton 'elliptic', $-v_2$, flow  data from the
FOPI Collaboration
(black dots with error bars) together with simulations using IQMD with a
stiff version of the EOS (HM, red) and a soft version (SM, blue), 
see Fig.~\ref{figEOS} for details to which we shall come back later.
The reaction is Au+Au at an incident energy of $1.2A$ GeV.
In \cite{reisdorf12} many more such data spanning the incident beam energy range
$0.15-1.5A$ GeV and varying centrality and system composition and size are
shown.

By plotting $-v_2$ (rather than $+v_2$) one sees in this energy regime an
enhancement of 
strength around mid-rapidity, i.e. predominantly
out-of-plane emission, which was given the suggestive name 'squeeze-out'
in pioneering, and predictive, theoretical work \cite{stoecker82} on the
subject.
Nowadays this phenomenon is named elliptic flow avoiding to conclude on the 
importance of shadowing and density gradients. 
At these energies the elliptic flow changes sign at high $|y_0|$
becoming predominantly in-plane.

As in  earlier FOPI work, collision centrality selection was obtained by binning 
distributions of  the ratio of total transverse to
longitudinal kinetic energies in the center-of-mass system, {\it ERAT}.
We estimate  the impact parameter $b$
from the measured differential cross sections for the {\it ERAT} 
using a geometrical sharp-cut approximation.
We wish to stress here that we have used the same procedure (ERAT)
for the simulations: the 
impact parameter, not accessible directly by experiment, does not enter the
analysis of the theoretical events.
A similar procedure was used to
correct for finite resolution of the reaction plane orientation(
see \cite{reisdorf12}).

We characterize the 
centrality by the interval of
the scaled 'impact parameter' $b_0$ defined by $b_0=b/b_{max}$,
taking $b_{max} = 1.15 (A_{P}^{1/3} + A_{T}^{1/3})$~fm.
The centrality chosen for the figure is $0.25<b_0<0.45$.
The choice of an optimum centrality for this study was guided
by the intention to compromise between the need for  sufficiently
sized 'shadow bars', composed by the spectator nucleons, on one side and
the requirement of sufficient compression in the participant volume, a
condition demanding a maximum size (maximum centrality) of the participant
matter.

Taking a closer look at $-v_2(y_0)$ we see that the predicted shape
is sensitive to the EoS {\it in the full rapidity range}.
To take advantage of this feature we introduce a quantity dubbed $v_{2n}$
defined by $v_{2n}=|v_{20}|+|v_{22}|$ where the parameters are fixed by a fit
to the flow data using $v_2(y_0)=v_{20}+v_{22}\cdot y_0^2$ in the scaled rapidity
range $|y_0|<0.8$. 
Such fits to the simulation data  
(here the error bars are of statistical
origin) are shown in the left panel of 
Fig.~\ref{v2} and then confronted
with the experimental data (the error bars are now dominated by systematic
errors) in the right panel.
While $v_{20}$ alone characterizes the flow at mid-rapidity only,
and the algebraic addition $v_{20}+v_{22}$ determines the flow at $y_0=1$
the quantity $v_{2n}$ combines features at both rapidities.  
The centrality is for $0.25<b_0<0.45$ and low momenta are
cut off both in the data and the simulations.
The cutting out of low transverse momenta, or equivalently, of scaled
 transverse 4-velocities ($u_{t0} < 0.4$),
originally forced by apparatus limitations, actually turns out to raise
the sensitivity of $v_2$ to the EOS as flow is generally converging to
zero at low momenta (see e.g. \cite{reisdorf12}).

\begin{figure}[t]
\centering
\includegraphics[width=0.85\textwidth]{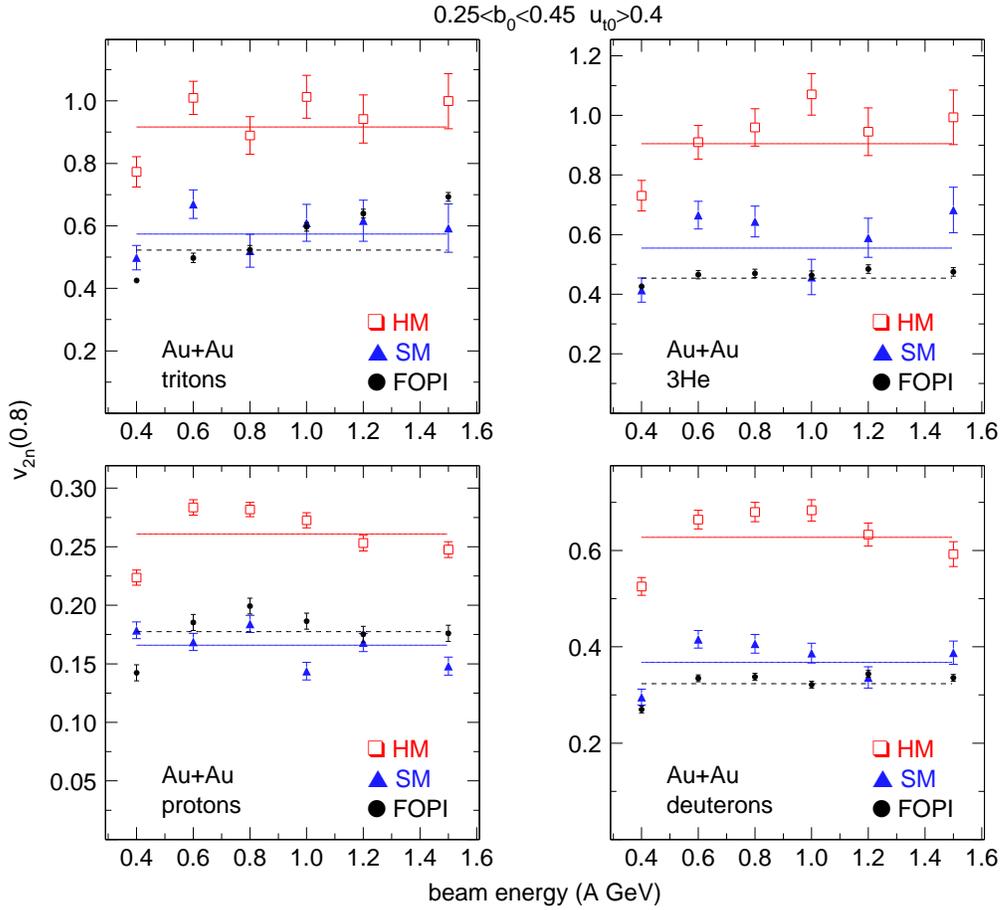}
\caption{Elliptic flow $v_{2n}$ for protons, deuterons, tritons, $^3$He as
  function of incident beam energy.}
\label{v2n}
\end{figure}

\begin{figure}[t]
\centering
\includegraphics[width=0.85\textwidth]{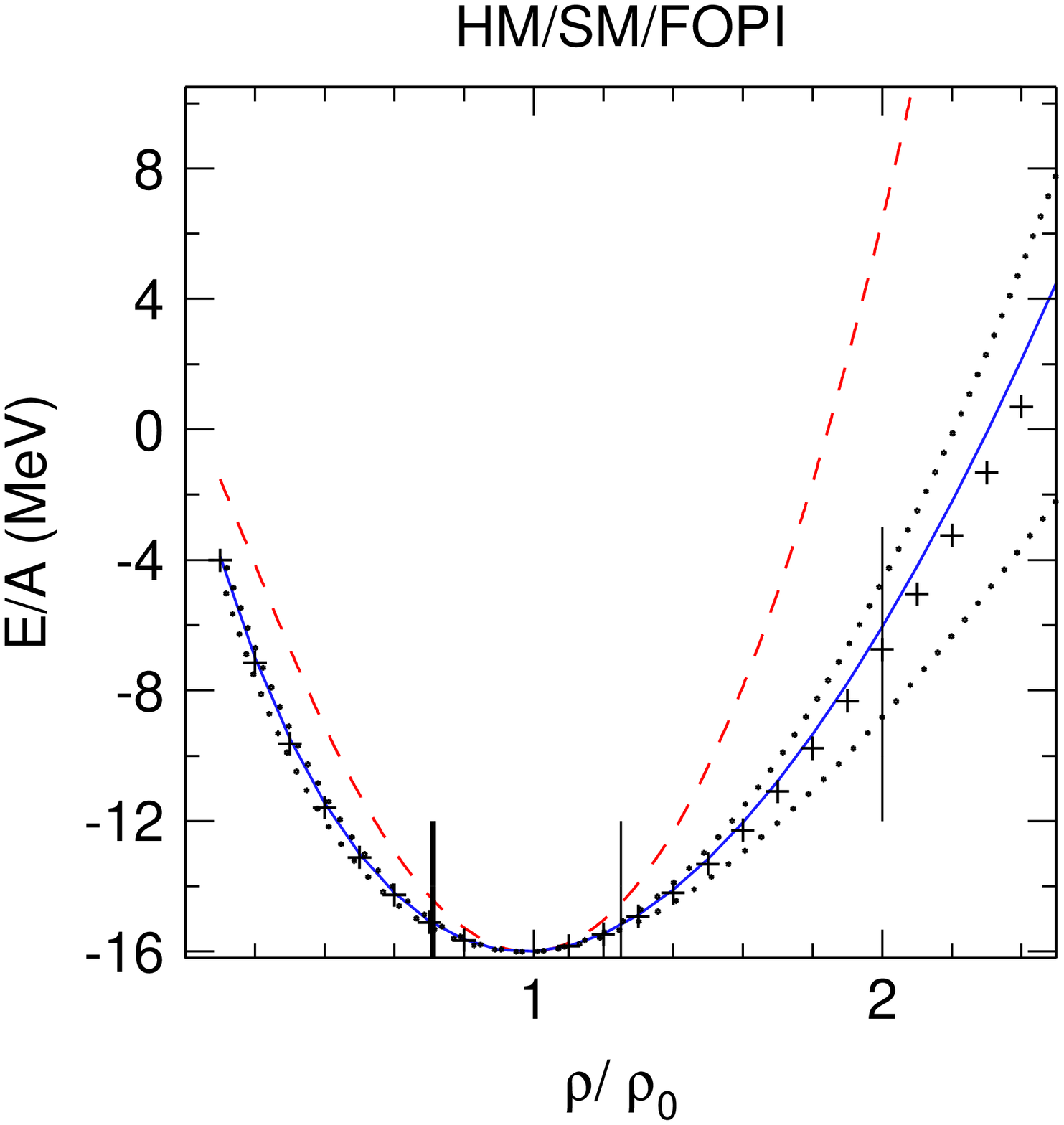}
\caption{Various symmetric nuclear matter EOS. Dashed (red) HM, 
full (blue) curve, SM. The dotted curves and the thin vertical bars 
delimit the FOPI EOS (crosses) constraints. The thick vertical bar is the
density relevant for the GMR according to \cite{khan13}.}
\label{figEOS}
\end{figure}

The $v_{2n}$ obtained for Au+Au between $0.4A$ and $1.5A$ GeV are shown in 
Fig.~\ref{v2n}
for protons (lower left) and deuterons (lower right), tritons (upper left)
and $^3$He.
As the beam energy dependencies are rather weak, we indicate the average
behavior by straight lines. The comparison of the data for $v_{2n}$ with
the calculations shows a rather convincing preference for SM.
The sensitivity is large: there is a factor $1.63\pm 0.06$ between HM and SM, a
difference exceeding significantly the indicated experimental error bars.
If we compare this factor with the fluctuations of the experimental data
points around the average values (dashed lines) we can estimate the
uncertainty of the deduced EOS.

To obtain quantitative conclusions on the EOS characterizer $K_0$ 
 for the energy range investigated in the present work 
 we have constructed Table I
which contains the relevant numbers (read off from the figure) with error bars.

\begin{table}
\caption{Derivation of the effective K using the elliptic flow observable 
$v_{2n}$}
\label{tab1}
\begin{center}
\begin{tabular}{llllllll}
\hline
&$v_{2n}$ SM & $v_{2n}$ HM & $v_{2n}$ FOPI & HM/SM & HM-SM & FOPI-SM & 
$\Delta K$ MeV  \\ 
\hline
$^1$H & $0.1658\pm 0.0030$ & $0.2609\pm 0.0027$ & $0.1774\pm 0.0028$ &
 $1.57\pm 0.07$ & $0.0951$ & $+0.0116$ & $+22\pm 8$\\
$^2$H & $0.3676\pm 0.0080$ & $0.6274\pm 0.0087$ & $0.3237\pm 0.0029$ &
 $1.71\pm 0.08$ & $0.2598$ & $-0.0439$ & $-30\pm 8$\\
$^3$H & $0.5740\pm 0.0214$ & $0.9161\pm 0.0252$ & $0.5223\pm 0.0048$ &
 $1.60\pm 0.08$ & $0.3421$ & $-0.0517$ & $-27\pm 17$\\
$^3$He & $0.5540\pm 0.0217$ & $0.9048\pm 0.0265$ & $0.4537\pm 0.0050$ &
 $1.63\pm 0.16$ & $0.3501$ & $-0.1010$ & $-52\pm 18$\\
\hline
\end{tabular}
\end{center}
\end{table}

The Table lists the average (straight lines in the figure) $v_{2n}$
obtained, in order, for SM, HM and FOPI, the ratio between HM and SM,
the difference between HM and SM, the difference between FOPI and SM, and
in the last column the modification of K (for SM) needed to account
for the difference FOPI-SM, assuming for simplicity
 a linear interpolation between the $K_0$ for HM and SM.
From the information using four different particles we thus obtain
a weighted average $K_0=190\pm 30$, while the theoretical ratio between HM
and SM is $1.63\pm0.06$. 
The used procedure is somewhat naive, but this is deemed uncritical
considering the adopted error on $K_0$. 

We have plotted in Fig. \ref{figEOS} the resulting EOS with its uncertainty
 band.
For comparison we have also plotted the 'trial' EOS, HM and SM, used in the
simulations. 
For the valid density range, indicated by vertical bars, see Section 
\ref{Simulations}.
Note the phenomenological EOS HM and SM include the saturation point
at $\rho/\rho_0=1$, $E/A=-16$ MeV by construction.
This fixes the absolute position of the curves: the heavy ion data
are only sensitve to the shape, i.e. the pressure which is
essentially the derivative.
Therefore the uncertainty of this fundamental point, about 0.5 MeV for
the binding energy  and 0.1 fm for $\rho_0$, is {\it not} included
in the uncertainty band. This will be important later when comparing to
some representative theoretical EOS, section \ref{Comparison}.
However, we can already at this stage conclude, in complete agreement
with ref. \cite{danielewicz02} that a stiff EOS, characterized by
$K_0=380$ MeV is not in agreement with flow data in the incident energy range
$(0.4 - 1.5)A$ GeV. For $0.4A$ GeV this had also been suggested in
\cite{stoicea04}.

\section{Simulations: the scenario}\label{Simulations}

\begin{figure}[t]
\centering
\includegraphics[width=0.9\textwidth]{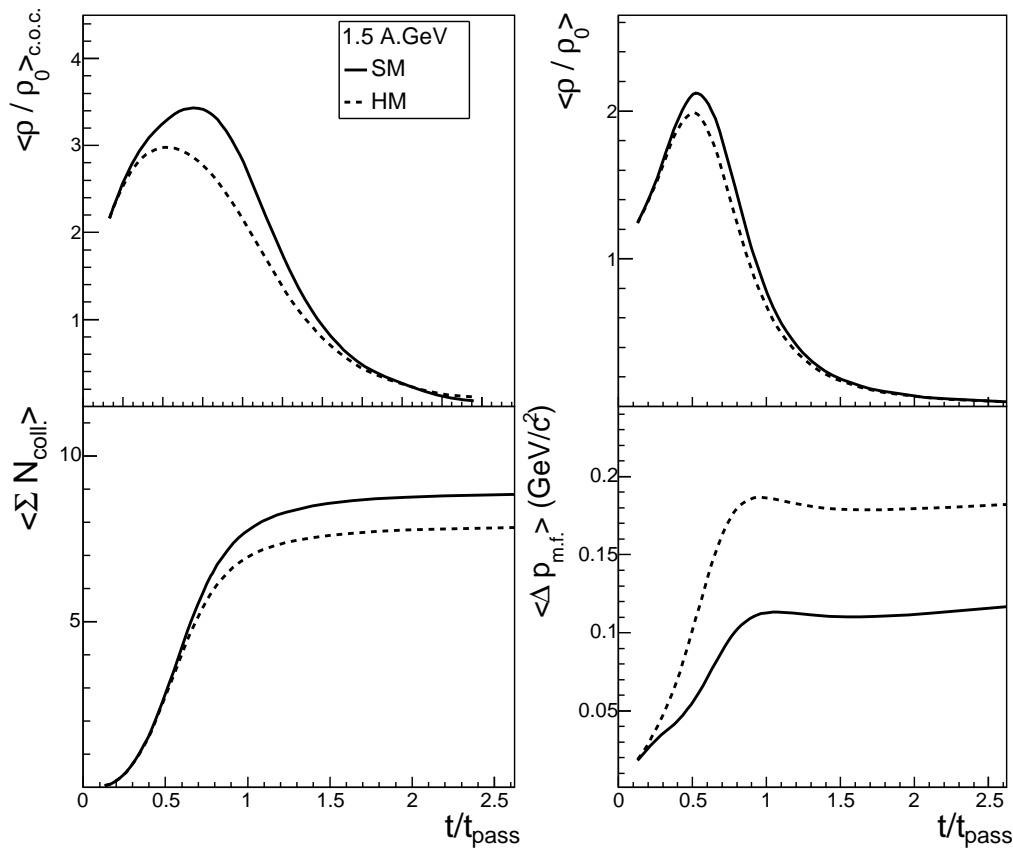}
\caption{IQMD predictions of the time evolution of various quantities in Au+Au
  collisions at 1.5A~GeV incident energy and an impact parameter $b=3~fm$ for
  various parameterizations of the EOS: soft EOS (SM) and hard EOS (HM),
  distinguishable through different line styles. The various quantities from
  left top to right bottom panels are: the reduced density probed by protons
  in the center of the colliding system, {\it idem} for all protons finally in
  the phase space region defined by $|y_0| < 0.8$ and $u_{t0} > 0.4$, the
  integrated (since time zero) sum of collisions experienced by these protons and
  their integrated momentum change due to the mean field (force).
}
\label{figALF1}
\end{figure}

\begin{figure}[t]
\centering
\includegraphics[width=0.45\textwidth]{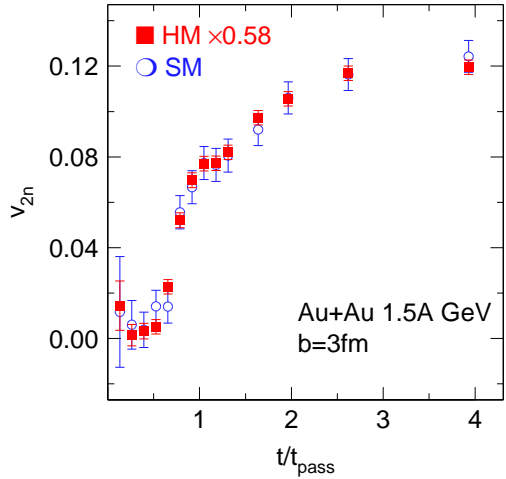}
\caption{
  IQMD results for $v_{2n}$ (see Section \ref{Analysis}) as a function
  of scaled time $t/t_{pass}$ for different EOS. The values of $v_{2n}$
  calculated with a hard EOS have been multiplied by a factor of 0.58 (filled squares) 
  to show the similarity with the shape of the SM predictions (open circles). 
  One observes that the ratio of $v_{2n}$ versus time reflects closely the ratio
  of the integrated momentum changes due to the mean field gradients
  shown in Fig.4. 
}
\label{figALF3b}
\end{figure}

\begin{figure}[t]
\centering
\includegraphics[width=0.45\textwidth]{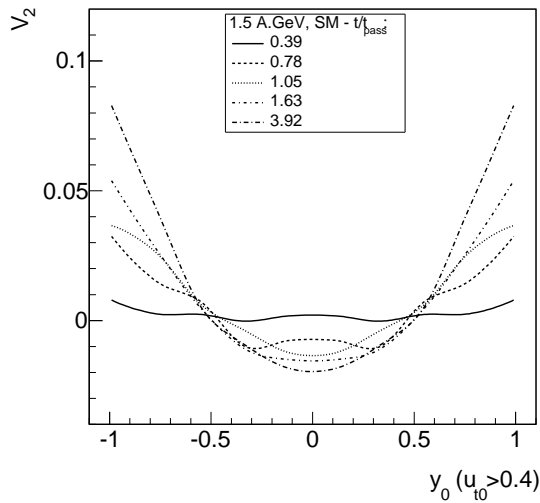}
\caption{Results of IQMD calculations for the elliptic flow $v_{2}$ of
  protons as a function of scaled rapidity $y_0$ at various values of the
  scaled time $t/t_{pass}$ (different line styles) and for an incident energy
  $E_{beam}=$1.5~A.GeV. The protons are selected
  to have high transverse momenta $u_{t0} > 0.4$ (see
  Section~\ref{Analysis}). The IQMD calculations are the same as in
  Fig.~\ref{figALF1} for a SM EOS.
}
\label{figALF3a}
\end{figure}


\begin{figure}[t]
\centering
\includegraphics[width=0.5\textwidth]{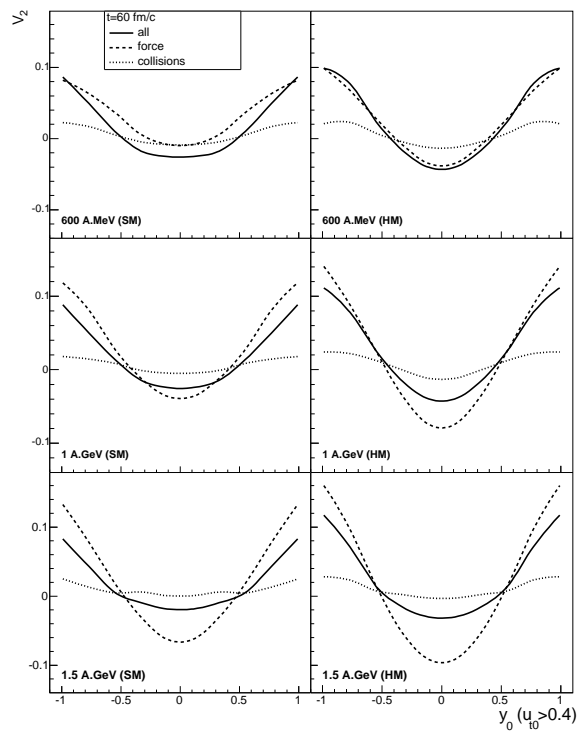}
\caption{Elliptic flow $v_{2}$ of protons having finally a high transverse momentum 
$u_{t0} > 0.4$ (see Section~\ref{Analysis}) as a function of their scaled
  rapidity $y_0$ predicted by IQMD.
Results for SM and HM EOS's are shown at 0.6, 1 and 1.5A~GeV incident energies 
at the late time $t=60$~fm/c independent of energy.
The total elliptic flow ('all', black lines) is presented along with its decomposition 
into the two contributions, i.e. mean field ('force', dashed lines) and 
collisions (dotted lines).   
}
\label{figALF5}
\end{figure}


\begin{figure}[t]
\begin{tabular}{lr}
\includegraphics[width=0.5\textwidth]{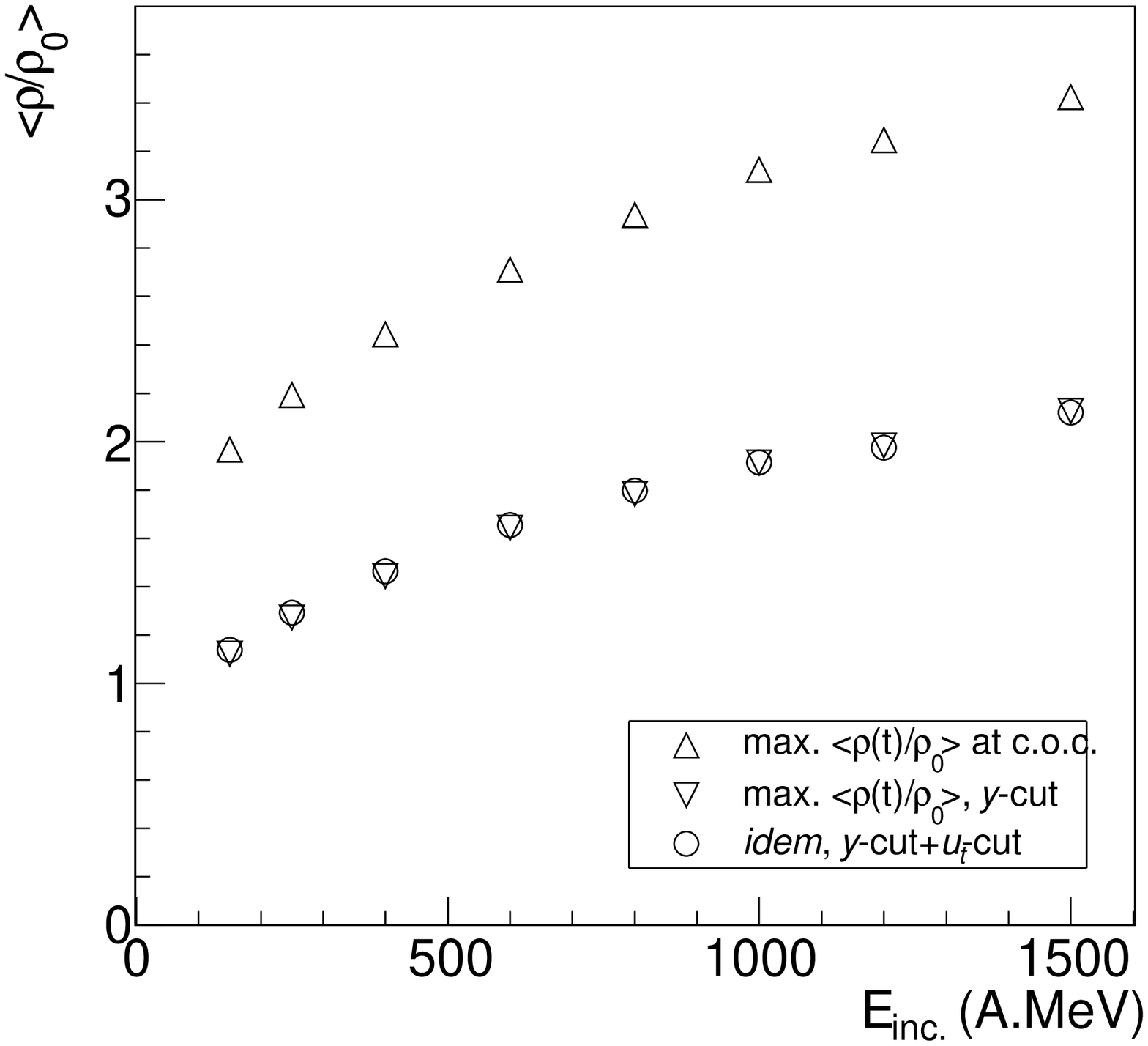}
&
\includegraphics[width=0.5\textwidth]{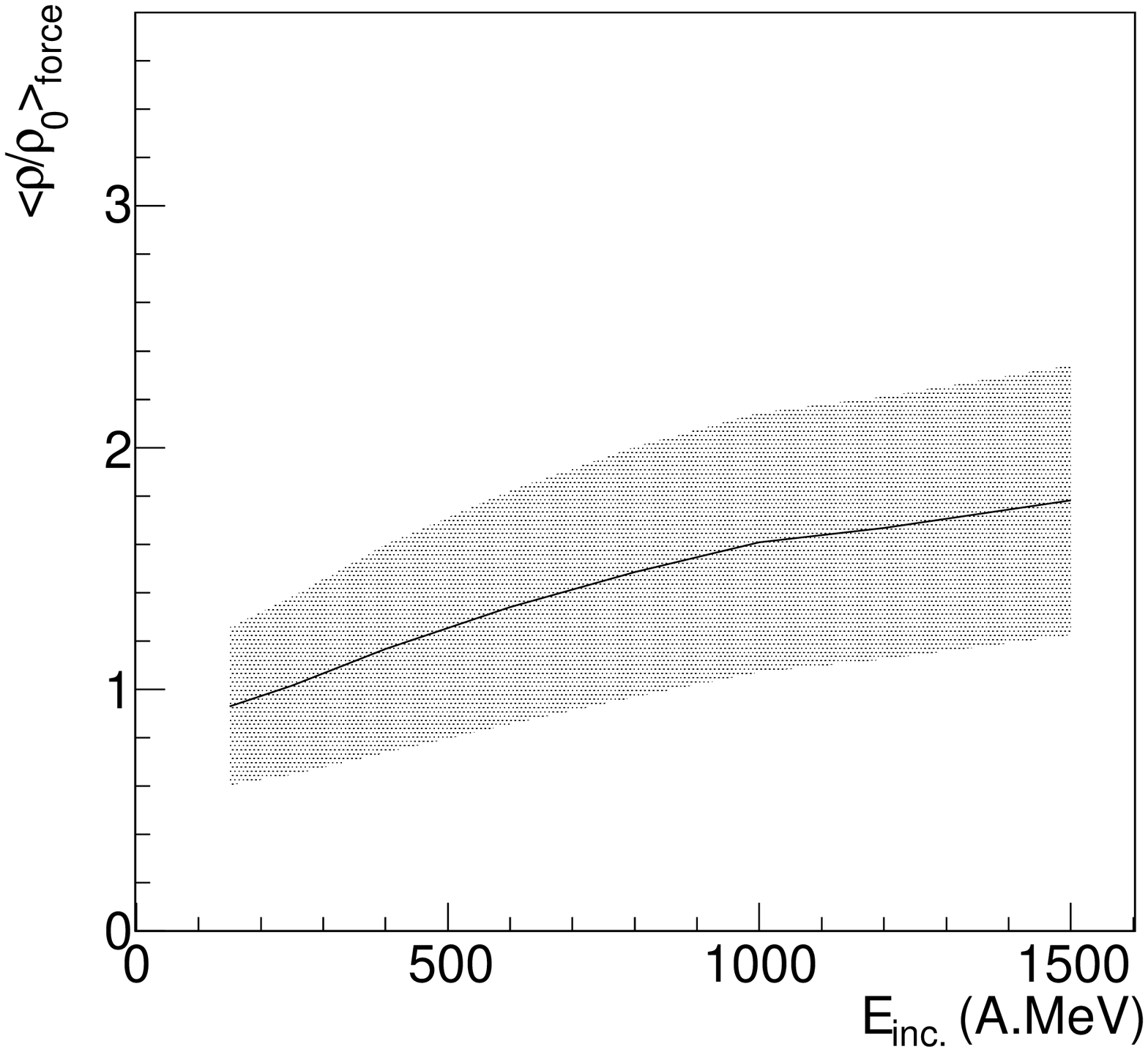}
\\
\end{tabular}
\caption{
Left side: The maximum of the density $<\rho(t)/\rho>_{max}$ reached in the central
volume of the collision (shown in  Fig.~\ref{figALF1} for 1.5A~GeV) as a
function of beam energy is shown as open triangles, the maximum of the density
$<\rho(t)/\rho_0>_{y-cut}$ for protons finally in the rapidity range
$|y_0|<0.8|$ as reversed triangles and as open circles the maximum of
$<\rho(t)/\rho_0>_{y-ut-cut}$ seen by protons with $|y_0|<0.8|$ and
$u_{t0}>0.4$. Right side: 
"Force-weighted" mean value of the reduced density (see text) seen 
by protons $u_{t0}>0.4$ and $|y_{0}|<0.8$ in their final configuration
averaged until the passing time. 
The error bars represent the standard deviations of the distributions. 
}
\label{figALF7}
\end{figure}

In order to determine the EOS empirically, it is reasonable to
start from low densities/pressures/temperatures and gradually raise 
these characteristics using sufficiently close steps.
This is necessary to 
become aware of discontinuities (even modest ones) possibly indicating
important structural changes.

The first step is to constrain the relevant density range. 
Isoscalar giant monopole resonance
(GMR) data ('breathing modes') have been used \cite{blaizot80} with the claim  to
determine the incompressibility around saturation density ($\rho=\rho_0=0.16$
fm $^{-3}$) where, by definition, the pressure in infinite symmetric nuclear
matter is zero.
An incompressibility $K(\rho_0)=K_0$ of $210\pm 30$ MeV was deduced in
 \cite{blaizot80} from the monopole energies.
 
In this context, it is interesting that recently some new insights have come
up concerning the $K_0$ value. In \cite{khan13} the authors have 
pointed out that
the GMR actually tests compressibility at about $2/3 \rho_0$ which is the
{\it average} density of large nuclei whose
finite depth surface area is not negligible.
The 'microscopic' theories using energy density functionals (EDF), adjusted to
reproduce the GMR data, then actually have to extrapolate to $\rho_0$
leading according to \cite{khan13} to $K_0=230 \pm 40$ MeV, the error
reflecting fluctuations between different EDF.

One should not expect that the value for the incompressibility of nuclear
matter is directly comparable
to the one used to characterize the tentative EOS adjusted to reproduce
heavy ion flows with transport codes, since the latter, as we will show, 
refer to
data associated to densities significantly larger than $\rho_0$, and there is no
basic principle which requires that the EOS is of simple parabolic shape over large
ranges of density. 

Whereas the determination of the nuclear matter EOS with GMR data is an
approach to saturation density 'from below',
our extensive IQMD simulations of the observations discussed in the
present work suggest that in the FOPI experiments, 
we are advancing towards saturation density 'from above'.

In contrast to model calculations done in an effort to reproduce the data,
the simulations discussed in this section are done at a single
impact parameter, since our goal is to elucidate the reaction scenario
and to characterize which 'typical' densities are probed
with the proton elliptic flow that has been extracted from the FOPI experiments. 
For this purpose one has to determine at which times -- 
in the course of the collision -- and medium conditions are linked with and influence the most the 
development of the elliptic flow.
Here, we restrict ourselves to the study 
of the $^{197}$Au+$^{197}$Au system at an 'intermediate' impact parameter $b=3 fm$. It 
is close to the 
upper limit of the centrality range chosen for the experimental data in Figs.~\ref{v2} and \ref{v2n}. 

In Fig.~\ref{figALF1} various quantities are plotted as a function 
of time scaled to the passing time $t/t_{pass}$, the time the nuclei pass each
other is approximated by $t_{pass}=2R/u_p$, where $u_p$ was introduced in
section \ref{Analysis}. The time zero is defined as  the 
the moment projectile and target nucleons have the {\it{first hard collision}}.
Typical values of $t_{pass}$ for various incident energies are listed
in Table~\ref{tabPassingTimes}.
The top panel of Fig.~\ref{figALF1} shows on the left hand side the
normalized density, $\rho/\rho_0$ (where $\rho_0$ is the nuclear matter ground
state density) seen by the protons in the central region of the collision. The
central region is defined as a sphere with a radius of 1 fm around the center.
This data is shown for reference and indicates the maximum density reached in
those reactions. For a soft EOS higher densities are reached than for a hard
EOS because the acting potential is less repulsive.
The upper right panel depicts the mean density seen by all protons with 
transverse 4-velocities $u_{t0}>0.4$ and $|y_0|<0.8$ in the final stage of the 
reaction. This cut was adopted in the analysis of the FOPI data and we
note that this selection removes most of the spectator particles.
One observes that the maximum of density probed by these protons in the course of the 
collision is reached at around the full overlap of the incoming system 
($t/t_{pass}=0.5$), and that the 'soft' EOS takes a slightly longer time 
to reach the maximum density, and reaches higher densities, 
as regard to the 'hard' EOS. 
The mean time-integrated number of collisions for selected protons is shown 
in the lower left panel, and in the right panel their mean integrated momentum transfer
caused by the potential.  
Obviously nearly all collisions have taken place until the passing time
and most of the momentum transfer due to the potential has been achieved, 
i.e. the influence of the potential becomes very small because the system is getting diluted. 
We only see a slight dependence on the stiffness of the EOS in the number of collisions,
caused by the different mean free paths of nucleons due to the different densities reached.
However, the momentum transfer caused by the
potential in the 'hard' case is much larger than in the 'soft' one, which reflects of the 
higher repulsion of the hard equation of state. 
Similar conclusions can be drawn for all incident 
energies that we have simulated (0.15 to 1.5~A.GeV). A summary of the maximum
densities reached during the time evolution of the reaction for a range of beam
energies is shown in Fig.~\ref{figALF7}.

\begin{table}
\caption{Passing time (see text) of the projectile and the target
 as a function of the incident energy in the $^{197}Au+^{197}Au$ collisions}
\label{tabPassingTimes}
\begin{center}
\begin{tabular}{lllllllll}
\hline
$E_{inc.}$ (A.MeV) & $150$ & $250$ & $400$ &
 $600$ & $800$ & $1000$ & $1200$ & $1500$\\
$t_{pass}$ (fm/c) & $48.3$ & $37.4$ & $29.6$ & 
$24.1$ & $20.9$ & $18.7$ & $17.1$ & $15.3$\\
\hline
\end{tabular}
\end{center}
\end{table}

Fig.~\ref{figALF3b} synthesizes  the dependence of
the variable $v_{2n}$ as a function of reduced time $t/t_{pass}$. This
observable saturates only
after two to three times the passing time with identical dependence for soft and hard EOS.
To underline the similarity we applied a scaling factor of 0.58 to the data for 
the hard EOS in Fig.~\ref{figALF3b}. Despite the fact that the 
absolute size of the signal differs by almost a factor of 2 for the EOS under considerations,
this difference becomes stable fairly quickly in the course of the reaction.
The elliptic flow signals $v_{n}$ themselves need rather long times to  obtain their
final (observed) shape. This is  illustrated in more detail in Fig.~\ref{figALF3a}:
the value of $v_2$ at mid-rapidity stabilizes quite fast, already 
at the full passing time,
but it needs nearly twice as long to reach its final shape
close to the target and projectile rapidities.  


We demonstrated that the mid-rapidity signal of $v_2(y_0)$ develops rather
rapidly within the IQMD model, stabilizing already after the passing
time. This is an indicator that the spectator clock
is working as anticipated. The elliptic flow at target and projectile rapidities 
takes a longer time to evolve since it originates also from late interactions with the spectator
material. 

A decomposition of the elliptic flow into potential and collisional 
contributions will elucidate the underlying reaction mechanisms.
In Fig.~\ref{figALF5} it is illustrated how the final $v_2(y_0)$ is obtained by 
NN collisions only, the mean field only, and by both operating together. 
The IQMD model allows to record the moment change due to collisions  and due 
to potential separately for each time step. Integrating all contributions over the
reaction time will yield the total momentum transfer vector due to the potential and 
analogously caused by the collisions. Both quantities have been analyzed to obtain
$v_2(y_0)$. Note, that the values of those different $v_2(y_0)$ contributions 
cannot be simply added, since they do not reflect an absolute magnitude
and the final momentum is obtained by a vector sum of the different contributions. 

We find the following features:

a) 
The collisional contribution to $v_2$ alone is small and little affected by different EOS.
This suggests that moderate medium changes of the NN cross sections
that go beyond 'zero order' Pauli blocking corrections would not critically affect
our conclusions on the EOS. Furthermore, the collisions by themselves cannot 
predict the measured elliptic flow. This illustrates the inadequacy 
of the cascade approaches for flow observables at these beam energies. 


b) The elliptic flow signal $v_2$ is more affected by the potential effects. One observes 
significantly differences of the $v_2(y_0)$ when applying different equations of states. 


Following this observation - that the elliptic flow signal is dominated by
the acting potentials - we propose to construct a
characteristic density which is connected to the elliptic flow of protons 
by calculating the mean value of the 
density seen by all protons with a final $u_{t0}>0.4$ 
{\it weighted by the strength of the force
due to the local mean field}. This quantity integrated up to the full passing time 
is displayed by the line in  
the right hand-side of Fig.~\ref{figALF7} as a function of bombarding energy. 
This variable is shown for the SM EOS which has been qualified by the FOPI measurements.
The shaded area represents the variances of the ''force-weighted'' densities
{\it (corresponding to finite size effects)}. 
Using such a method we are not limiting ourselves to some small most central volume
of the system 
a procedure which then naturally yields significantly larger densities. 
Those are depicted in the left hand-side of Fig.~\ref{figALF7}.
The maximum density in the central volume
$max <\rho/\rho_0>(t)$ is shown as function of bombarding energy, as well as the
maximum density seen by all protons with final $|y_0|<0.8$ and the maximum density
of protons with $|y_0|<0.8$ and $u_{t0}>0.4$. In the innermost center of the
collision densities up to 3.5$\rho_0$ are reached at the highest beam
energies measured by the FOPI collaboration. The density experienced by all protons
measured in the region around mid-rapidity $|y_0|<0.8$ is substantially lower
(up to 2$\rho_0$),and this value is not changed by the additional
experimental cut $u_{t0}>0.4$.  


Our method of averaging with the strength of the force 
is in the same spirit as proposed in \cite{khan13} for the GMR.
As can be seen in Fig.~\ref{figALF7}, the gap to the $2/3\rho_0$ data of the GMR could
be closed by lowering the incident energies at least down to
0.1~A.GeV. Even if the densities extracted here are smaller than the maximum density
reached during the collisions, the presence of the highly compressed matter
(up to 3$\rho_0$) is mandatory for building up the high density gradients
leading to the repulsion of particles and finally the elliptic flow signals.




\section{Comparison to microscopic calculations}\label{Comparison}

In view of the 
large number of proposed nuclear matter EOS in the
 literature, see for example the recent discussion of EOS derived using 240 (!)
'phenomenological' (Skyrme interactions) approaches in \cite{dutra12},
we cannot  here  compare our constrained 'experimental' EOS in
Fig. 3  with all these various predicted behaviors,
 but hope that future theoretical efforts will consider our results for
 the density range up to $\rho/\rho_0=2.5$. 

By 'microscopic'  calculations  we define here
theoretical efforts to calculate the infinite nuclear matter EOS
using nucleon-nucleon and pion-nucleon scattering data and some information on
multi-body forces (3N, 4N) from light nuclear clusters, including
of course the deuteron, {\it without introducing further ad hoc parameters}.
Taranto et al. \cite{taranto13} have recently tested five such models for
their consistency with known astrophysics and nuclear physics constraints.
They came up with open questions concerning the predictions for high densities
($\rho > 3\rho_0$) that one needs to know in compact star modeling.
An intriguing point was that three of the models violated causality at
the predicted central densities and the suggestion that even 
Dirac-Brueckner-Hartree-Fock (DBHF) calculations may be in error due to approximations made
in order to make the calculation tractable. 

We choose here to confront three {\it representative} microscopic calculations
with our new constraints.
First a DBHF calculation \cite{katayama13} of both symmetric and pure neutron
matter which was presented in a conveniently parameterized form
allowing to estimate with use of the parabolic assumption 
(involving $\delta^2$) the
deviation of the EOS in going from strictly symmetric to Au-like
($\delta^2=0.0391$) composition, where $\delta=(N-Z)/(N+Z)$.
The second comparison uses work from the 'Idaho group' \cite{sammarruca12}
allowing to compare the historically earlier DBHF approach with the more
recent chirally motivated theories.
Finally, we compare with two different EOS published in 2005 \cite{fritsch05}
one having the interesting feature of including the possibility of (virtual)
$\Delta$ excitations {\it explicitly}.
%

\begin{figure}[t]
\centering
\includegraphics[width=0.70\textwidth]{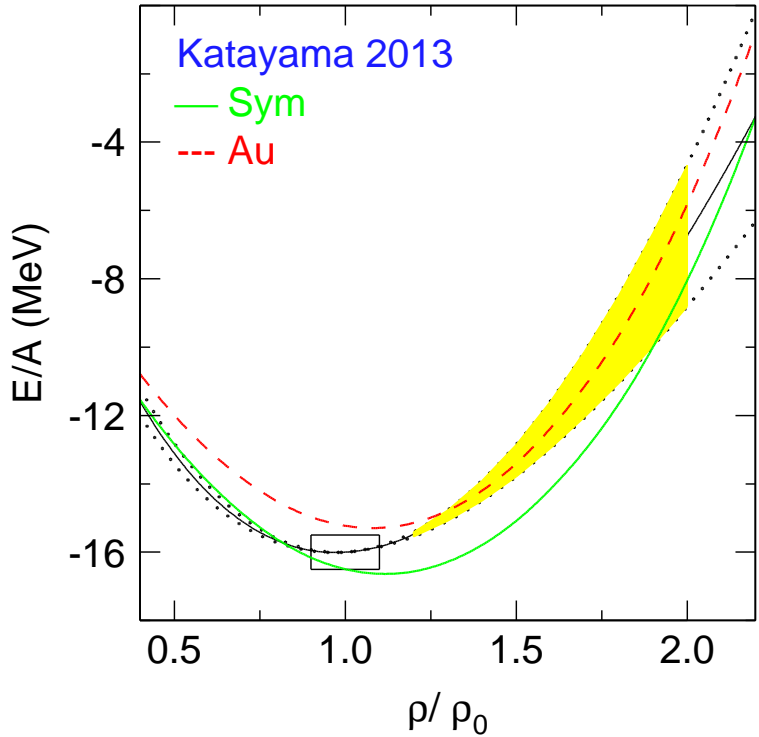}
\caption{Symmetric nuclear matter EOS from ref. \cite{katayama13} using
Bonn A potential  (full green curve). The dashed red curve is the EOS for
Au-like systems using neutron matter calculations from \cite{katayama13}.
The shaded error band surrounding the black curve is the FOPI-IQMD constraint.} 
\label{katayama}
\end{figure}

The EOS calculated with DBHF in \cite{katayama13} 
using the Bonn A \cite{brockmann90} nucleon-nucleon
potential (full green curve) is confronted in Fig. \ref{katayama} with our
constrained FOPI-IQMD EOS (shaded band around the median black curve).
The authors \cite{katayama13} have made efforts to check the validity
of averaging approximations often used in DBHF to reduce the
computational cost.

As mentioned before we have also estimated the 'correction' to the
symmetric EOS due to the degree of isospin asymmetry in gold (Au),
see the dashed curve.
It is located roughly $1-1.5A$ MeV above the symmetric system.
It is visible that the shape of the 'Au' EOS is not dramatically
different from the strictly symmetric matter EOS.
Of course the heavy ion flow data, being sensitive only to the pressure,
i.e. essentially the derivative of such curves, does not allow to
determine the absolute vertical position of the energy/nucleon EOS.
The uncertainty of this position is roughly indicated by the rectangle
in the figure. 
As there seems no clear consensus on the widths of this 'saturation rectangle'
in the literature we have adopted here approximately the sizes given in two
theoretical papers, \cite{furnstahl13,hebeler11}
If one adopts conclusions from given observables (such as nuclear
masses and electron-scattering data) one can come to tighter, but still 
model dependent limits.
As just one example, 
one obtains very tight limits ($<$0.2~MeV) on the bulk
binding energy if one interprets the mass-number proportional coefficient
 as holding
for infinite NM, see a recent effort in \cite{moeller12}.

The narrowing 'error' band of the empirical curve in approaching the
saturation point is a consequence of 
constraining the curve to cross the most probable saturation value
and ignores the shown uncertainty of this very fundamental point.

Constraining the stiffness of pure neutron matter at densities
significantly exceeding the saturation density is and will be a difficult
task as realistic heavy ion systems, even at radioactive beam facilities,
will not exceed $\delta^2$ values larger than $5\%$.
It is therefore not surprising that information relevant to compact star
physics \cite{tsang12} is so far primarily limited to subsaturation densities
that are accessible by other means.

\begin{figure}[t]
\centering
\includegraphics[width=0.70\textwidth]{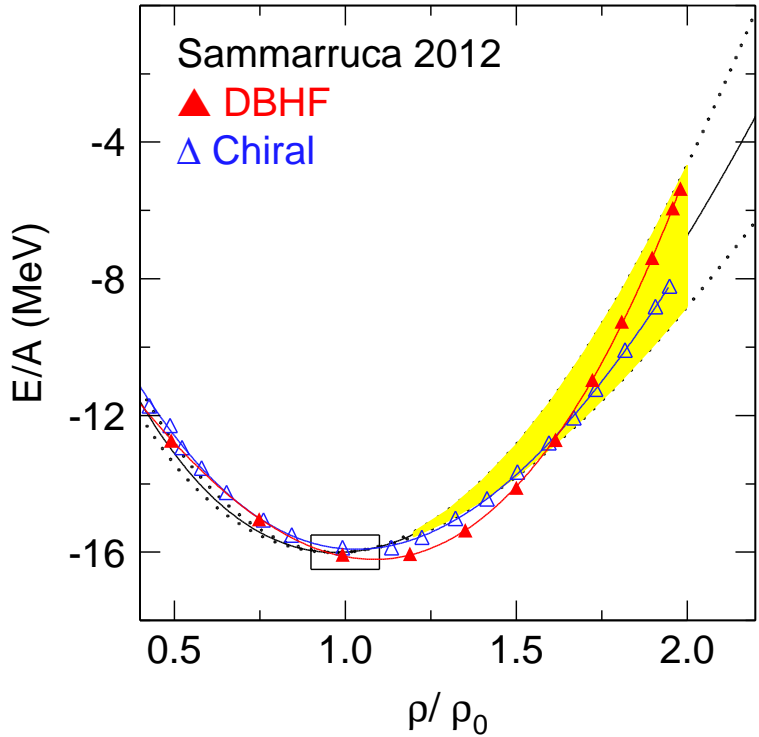}
\caption{Symmetric nuclear matter EOS after ref. \cite{sammarruca12}.
Full (red) triangles: DBHF calculation, open (blue) triangles:
chiral effective field theory. 
The shaded band is the FOPI-IQMD constraint.} 
\label{sammarruca}
\end{figure}

In Fig. \ref{sammarruca} we show two symmetric nuclear matter EOS from
ref. \cite{sammarruca12}. 
One, full triangles, labeled 'DBHF' in the figure,
is using a meson theoretic potential together with
the DBHF method, the other, open triangles, labeled 'Chiral',
makes use of effective field
theory (EFT) with density dependent interactions derived from leading order
chiral three-nucleon forces.

One can see that the DBHF curve tends to become stiffer at densities exceeding
$\rho > 1.5\rho_0$. On the other hand the 'chiral' version seems to match
close to perfectly with the FOPI-IQMD constraint (shaded).
As the EFT approach, respecting QCD symmetries, is a low-momentum theory
sorting the contributions of various order according to powers of momentum
the question arises up to what density one can trust the predictions.
In \cite{machleidt07} Machleidt pointed out that chiral NN potentials do
not make any reasonable predictions beyond 300 MeV laboratory energy.
He ends up concluding that 'one may trust the chiral perturbation theory up
to densities around $4\rho_0$'.

\begin{figure}[t]
\centering
\includegraphics[width=0.70\textwidth]{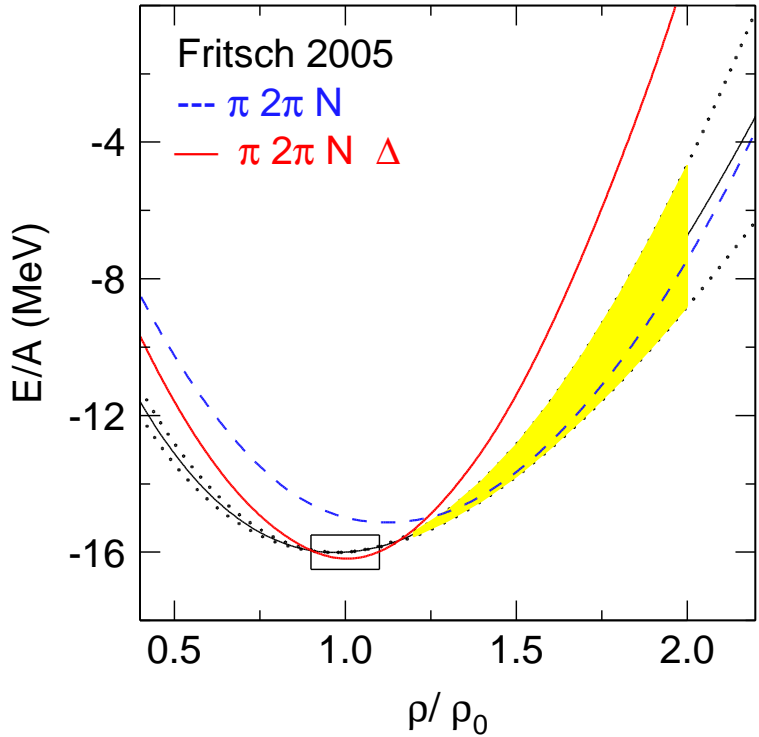}
\caption{Symmetric nuclear matter EOS after ref. \cite{fritsch05}.
The dashed (blue) EOS curve includes pions and nucleons as active degrees
of freedom.
The full (red) curve includes effects from $2\pi$ exchange with virtual
$\Delta$-excitation. 
The shaded band is the FOPI-IQMD constraint.} 
\label{fritsch}
\end{figure}

The treatment of the short range parts of the nucleon-nucleon forces outside
of QCD lattice theory is still subject to different approaches, not only
between the above mentioned DBHF and Chiral EOS, but also in some other
'chiral approaches to nuclear matter' \cite{fritsch05}.
As we wish to limit the scope and size of the present contribution, we will
not describe or comment these various approaches.
However we can show, see Fig. \ref{fritsch}, how the theoretical
investigations may use the FOPI-IQMD constraint as rough guide line.

Using the chiral approach, the authors \cite{fritsch05} predict
two rather different EOS depending on whether they include or not
virtual $\Delta$ excitations. See the figure caption.
The predictions including the $\Delta$ are in agreement with the empirical
saturation point.
The theory is not completely free of some adjustments that go beyond the
use of $\pi N$ and $NN$ scattering data.
The short-range parameters of the theory have been 'fine-tuned' to
reproduce the $(-16)$ MeV binding energy per nucleon at saturation.
However, the virtual $\Delta$-excitations then help to locate the EOS at the
right horizontal place around $\rho=0.16$ fm$^{-3}$.
A look at Fig. \ref{fritsch} shows that the $\Delta$ leads to a rather
marked stiffening of the EOS ($K_0=304$ MeV) projecting the curve outside
the shaded FOPI-IQMD constraint.
It is not obvious to us what this discrepancy means.
It could be that the apparent 'heavy-ion EOS' actually differs from a
'cold' EOS because, due to the finite temperature in the reaction the
$\Delta$ are real rather than virtual.
The theoretical '$\Delta$ stiffness' could then be a dispersion effect
rapidly changing with temperature.
If so, it would be eminently important to pursue this question.
Naively, one would expect that the (real) possibility to store energy 
internally into the nucleon (at high T) would lower the kinetic energies and
hence the pressure and stiffness, some kind of mild 'phase transition' when
new degrees of freedom open.
\section{Discussion}\label{Discussion}

\begin{figure}[t]
\centering
\includegraphics[width=0.70\textwidth]{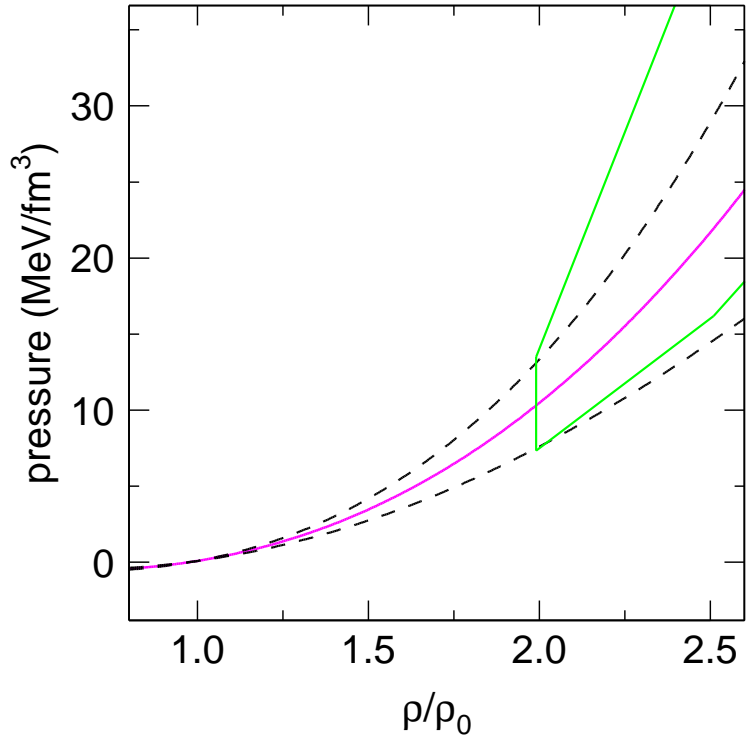}
\caption{Nuclear EOS in terms of pressure versus density. The pressure
as deduced from the soft EOS which was
used in the IQMD model calculations is shown by the magenta line. The dashed
curves enclose the error band in the determination of the relevant pressure
region. The area framed with lines originates from the analysis of \cite{danielewicz02}
}
\label{fig-eos-pc}
\end{figure}

We have shown in section \ref{Analysis} that a single parameter which
we dubbed $v_{2n}$,
characterizing elliptic flow over a large rapidity interval ($80\%$ of
the initial gap) allows to discriminate rather clearly between a soft (SM)
and a stiff (HM) EOS.
$v_{2n}$ differs by a factor 1.63 over an energy range extending from
$0.4A$ GeV to $1.5A$ GeV when using IQMD, while the experiment allows to
constrain $v_{2n}$ within roughly a factor 1.1.
The relevant density range was estimated from the simulations to span
$\rho_{max}=(1.1-3.0) \rho_0$, where the high densities reached during the
collisions are important for the evolution of the final elliptic flow
pattern. 
This makes the 'flow method' competitive and complementary with the
'kaon method': the latter reaches such a high SM/HM discrimination only 
around $0.8A$~GeV far below threshold.
It is gratifying that both methods lead to the same conclusion, however.

In Fig.~\ref{fig-eos-pc} the range of pressure and density assessed in this
work is shown by the area enclosed by the dashed lines. The pressure for
symmetric matter at zero temperature predicted by the FOPI-IQMD
constrained EOS with $\kappa = 190 \pm 30 MeV$
is shown as magenta line. To put these results in context with
others they are compared to the constraints reported in the analysis of \cite{danielewicz02} at higher
energies and respective densities. Our data is in agreement with the earlier
findings giving slightly more restrictive constraints towards a more soft nuclear matter
EOS at higher densities. 

Convincing conclusions on basic nuclear properties require a successful
simulation of the full set of experimental observables with the same code
using the same physical and technical parameters.
For a number of observables this goal has been reached, for some
other data this is not yet the case.

It is well known that in the beam axis system elliptic flow is not
independent of the flow axis and hence the magnitude of the directed flow.
In contrast to ref. \cite{danielewicz02}, the IQMD model reproduces the 
directed flow using the same EOS parameters.
In ref. \cite{reisdorf12} this is shown (Fig. 49) for directed flow of
deuterons in the same beam energy and centrality range.
Here too the experimental errors were significantly smaller than the
HM/SM difference.

Furthermore, in the same work, Fig. 51, radial flow of the light clusters
was well reproduced (and found to be insensitive to EOS).

Stopping information \cite{reisdorf04, reisdorf10} and Fig. 52 in \cite{reisdorf12}
 relevant to the question of in-medium nucleon-nucleon
cross sections, requires still some fine tuning on the $20\%$ level.
Incomplete stopping could mockup a soft EOS if it is incorrectly 
accounted for in the simulations \cite{fuchs03}.

Pion yields, measured with $10\%$ accuracy, \cite{reisdorf07}, differ only
by about $10\%$ between the HM and SM options, and the yields were found
to have about $20\%$ lower values than the SM option of IQMD.
It should be kept in mind that the elementary pion cross sections are
 known with limited precision.  

Some less desirable features of our version of IQMD will not be hidden:

a) the momentum dependencies of protonic $v_1$ are not correctly reproduced,
while it is increasingly correct the heavier the cluster;
this could be due to a lack of quantum fluctuations in a 
semiclassical code; we expect that fluctuations have a diminished
influence on $v_{2n}$ because it is determined by a large part of
phase space;

b) the degree of clusterization is underestimated; this deserves further
investigations.

 It is worthwhile mentioning that in \cite{reisdorf12} some
comparisons were also done with lower energy data.
At $0.15A$ GeV the difference HM/SM was found to be rather small.
The deuteron elliptic flow was well reproduced by IQMD, Fig. 38.
The small sensitivity to EOS at beam energies well below $0.4A$ GeV 
is not surprising in view of the flatness of the energy/nucleon EOS
(see Fig. 3) in this regime, implying relatively low pressures.
We also remark that at $0.25A$ GeV the $v_1$ flow of all light
clusters was well reproduced.
The reproduction of low energy data, where the effect of the EOS is still
moderate, is useful to ensure the adequacy of some technical parameters
of the code.

Last, but not least, essentially the  same IQMD code
(except for some kaon specific sections) was used 
for the 'kaon method'.
 
Despite these encouraging results,
there are two important points to be aware of:

1. data that are claimed to constrain such fundamental properties as is the
EOS of nuclear matter require confirmation by independent experimental efforts

2. there is an urgent need to prove that the conclusions using a particular
transport code, here IQMD, are not limited to using this particular code.

We shall briefly comment on these two points.
 
Before the 'experimental' EOS is finally ready for textbooks one needs a
general consensus on the measured data with use of various apparatus. 
At the time of writing this goal is not yet reached in a satisfactory way.
In ref.~\cite{danielewicz02} two of the at that time available experimental
 data points on elliptic flow within the
SIS/BEVALAC energy range (below 2$A$ GeV) survived the critical eye of the
authors: a point ($|v_2|\approx 7.7\%$)
 around 0.4$A$ GeV from the Plastic Ball Collaboration
\cite{gutbrod89} and a point near 1.1$A$ GeV ($6.2\%$) from the EOS 
Collaboration
both obtained at the BEVALAC accelerator in Berkeley (USA) with a different
 apparatus. 
We could not find the (non-rotated) Plastic Ball value in the literature
 as the authors of refs.
\cite{gutbrod89,gutbrod90} choose at the time the so called 'flow axis'
 as reference  to determine elliptic flow. 
To our knowledge the AGS data  \cite{pinkenburg99}
 were not 'rotated' 
(see \cite{reisdorf12} for a brief discussion of the issue of rotated
 reference systems).
A close comparison in \cite{reisdorf12} with the Plastic Ball's rotated
data suggested disagreement with the FOPI data in at least part of the
overlapping incident energy range.
The $1.1A$ GeV point accepted in \cite{danielewicz02} was first cited
in \cite{pinkenburg99}, but we could not find a detailed documentation in the
refereed literature on how possible apparatus cuts and other effects were
taken into account.
Optimistically, the $1.1A$ GeV value is compatible with the FOPI data if
one assumes a compensation of cut effects (present in the FOPI data)
and slightly different beam energies.

In \cite{reisdorf12} comparisons to proton data from KAOS \cite{brill96}
used by Danielewicz
\cite{danielewicz00} to tune momentum dependencies were also done.
The agreement was acceptable but not perfect, see \cite{reisdorf12} Fig. 35
for details.

The second important point to be discussed is that the conclusions drawn from
the comparison with simulation data obtained from various transport 
codes must be consistent.
There are several issues that need further efforts by the community:
momentum dependencies (including clean Lorentz covariance at beam energies
exceeding $1A$ GeV), clusterization and entropy balances, in-medium
nucleon-nucleon reactions etc. 
Efforts to bring transport code experts together, to see if a convergence of
analyses results can be reached, have been done \cite{transport} but
urgently need to be continued including energies up to FAIR accelerator
design.

What can we say about the future besides asking for efforts to
reach consensus on both data and transport code analyses?

The sensitivity of $v_2$ to the assumed stiffness of the  EOS
 was shown \cite{danielewicz02} to diminish rapidly below 0.4$A$ Gev and
 above 6$A$ GeV.
We now tend to associate the weak sensitivity at low energies to the
combined action of insufficient stopping \cite{reisdorf04} and
still relatively modest compression.
At energies well above 4$A$GeV, on the other hand, the shadowing effect
by spectator nucleons (and therefore the  time clock) is no longer
very effective
as suggested by the observed sign change of $v_2$ at 4$A$ GeV
 \cite{pinkenburg99}.
Presumably the spectator clock can be used  to
extend improved EOS constraints to densities (3-4 $\rho_0$) higher 
than was possible in the present work (limited to beams of 1.5$A$ GeV
incident energy) in future
accelerator systems such as FAIR in Darmstadt, Germany.
However it is clear that beyond 4A GeV, other ideas are needed to
extract EOS information from heavy ion data.


As a main conclusion of the present work we believe we can say that the
feasibility of establishing reasonably tight  empirical constraints on the
nuclear EOS has been demonstrated.
\section*{Acknowledgments}
This work was supported by the Helmholtz
International Center for FAIR within the framework of the LOEWE
program (Landesoffensive zur Entwicklung
Wissenschaftlich-\"Okonomischer Exzellenz) launched by the State
of Hesse and under the contract number 13-70 by the German-French Exchange 
Program supported by GSI and IN2P3. 





\end{document}